\documentclass{cfm2011}

\usepackage[T1]{fontenc}
\usepackage{amsfonts}
\usepackage{amssymb}
\usepackage{amsmath}
\usepackage[applemac]{inputenc}
\usepackage{graphicx}
\usepackage{color}

\title{Modélisation multidomaine du comportement magnéto-mécanique des aciers dual-phases}
\author{F.S. Mballa-Mballa\inst{a,b}, O. Hubert\inst{a}, S. Lazreg\inst{a}, P. Meilland\inst{b}}

\instlabel{a}{LMT-Cachan (ENS-Cachan/UMR CNRS 8535/UPMC/Pres Universud Paris) 61, avenue du président Wilson 94235  Cachan Cedex}
\instlabel{b}{ArcelorMittal Maizières Research BP 30320 - Voie Romaine F-57283 Maizières-lès-Metz Cedex}

\begin{document}

\maketitle

\begin{resume}
	 \textit{ La microstructure des aciers dual-phases et leur comportement mécanique sont fortement sensibles aux variations de procédé (traitements thermiques). Un contrôle en ligne par méthode magnétique est envisagé, la mesure s'appliquant à un matériau soumis à un état de contrainte. Le dual-phase est un milieu biphasé (ferrite/martensite), où chacune des phases est considérée comme étant une sphère immergée dans un milieu homogène équivalent. La modélisation de chaque phase s'appuie sur un modèle magnéto-mécanique couplé. Il s'agit d'un modèle monocristallin explicite représentatif du polycristal isotrope de la phase considérée. La mise en place de règles de localisation permet la simulation du milieu biphasé. Expériences et modèle sont comparés.}
\end{resume}

\begin{abstract}
    \textit{The microstructure and mechanical behavior of dual-phase steels are highly sensitive to the variation of the process (heat treatments). Online control by magnetic method is relevant. A measurement under applied stress must be considered. The dual-phase is a two-phase medium (ferrite / martensite). Each phase can be considered as a sphere embedded in a homogeneous equivalent medium. The model used for each phase is based on a magneto-mechanical coupled model. This is an explicit single crystalline model representative of the behavior of the corresponding phase. Localization rules allow the simulation of the two-phases medium. Experiments and modeling are compared.}
\end{abstract}


\section{Introduction}

\begin{figure}[htbp]
\begin{center}
\includegraphics[width=7cm]{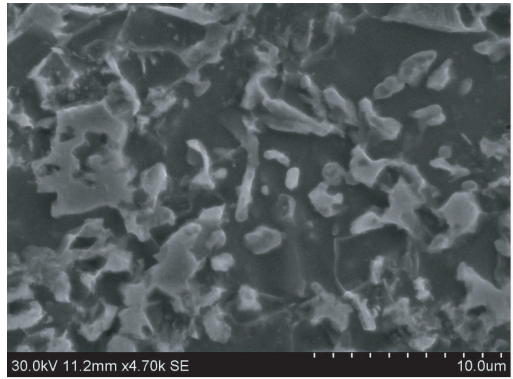}
\caption{\label{dp} Microstructure d'un acier dual-phase: martensite en blanc, ferrite en sombre.}
\end{center}
\end{figure}

Les dernières années ont vues un intérêt grandissant des industries automobiles pour l'utilisation d'aciers à haute performance tels que les aciers dual-phases (DP). La production de ces aciers implique plusieurs procédés de fabrication : métallurgie primaire et métallurgie secondaire, coulée, laminage (à chaud/froid) et traitements thermiques. Ces procédés conduisent à une microstructure biphasée principalement composée d'îlots de martensite dure dispersés dans une matrice ferritique ductile (figure \ref{dp}) en proportion variable selon l'histoire thermo-mécanique du matériau. Leur microstructure (fraction et composition des phases) étant fortement sensible aux variations du procédé de fabrication (traitement thermique, laminage), Les industriels cherchent à mettre en place une méthode de contrôle non destructive en ligne. La méthode employée exploite les propriétés magnétiques de ces matériaux. La mise en \oe uvre de ce procédé de contrôle rentre dans le cadre du projet ANR DPS-MMOD \cite{ANR}.

Le travail présenté dans cette communication a été réalisé dans le cadre de ce projet: il s'agit d'une contribution à la modélisation du comportement magnéto-mécanique des aciers DP. La méthode de modélisation retenue s'appuie sur le modèle multidomaine, développé au LMT-Cachan. Ce modèle est basé sur la partition en domaines magnétiques des matériaux magnétiques. Il s'agit d'un modèle monocristallin. On montre sous certaines hypothèses qu'il existe une direction de sollicitation pour laquelle le comportement obtenu est représentatif du polycristal isotrope du matériau concerné. Il s'agit, dans le cadre de l'application recherchée, de prendre en compte:

\begin{itemize}
 \item  l'orientation du chargement magnétique dans le repère du monocristal représentatif.
 \item  un chargement mécanique statique imposé et ses effets sur la mesure.
 \item  la nature biphasée de la microstructure de l'acier DP, ce qui suppose de mettre en place des règles de localisation mécanique et magnétique adaptées.
\end{itemize}

\section{Modélisation multidomaine \cite{lazreg1,lazreg2}}

\begin{figure}[htbp]
\begin{center}
\includegraphics[width=13cm]{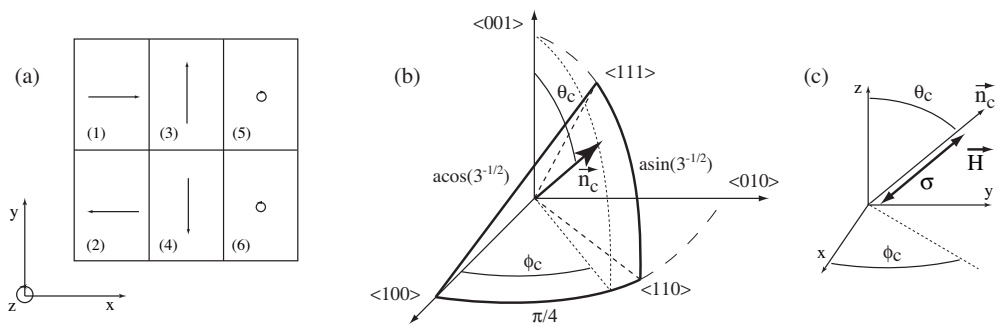}
\caption{\label{mmd} (a) représentation simplifiée du monocristal; (b-c) triangle standard et paramétrage du chargement.}
\end{center}
\end{figure}

Les deux phases en présence (ferrite $f$ / martensite $m$) sont ferromagnétiques: elles s'aimantent et se déforment en présence d'un champ magnétique. Les deux phases sont modélisées séparément à l'aide d'un modèle de comportement proposé récemment: le modèle multidomaine \cite{lazreg1}. Ce modèle prend son origine dans la configuration en domaines magnétiques des matériaux magnétiques. On considère un monocristal de fer de symétrie cubique divisé en 6 familles de domaines notés $\alpha$ (	axes $<100>$ du cristal) correspondant aux 6 directions de facile aimantation (figure \ref{mmd}a). On note $\vec{\gamma}^\alpha= (\gamma_1^\alpha\ \gamma_2^\alpha\ \gamma_2^\alpha\ )^T$ le vecteur directeur de chacun des domaines (\ref{eq0}). Il est colinéaire au vecteur aimantation $\vec{M}^\alpha=M_s.\vec{\gamma}^\alpha$, où $M_s$ désigne l'aimantation à saturation du matériau. Le vecteur aimantation est initialement confondu avec les axes du cristal. Sous l'effet d'un chargement magnétique ou mécanique, les fractions de domaines ainsi que les directions d'aimantation changent. 

\begin{equation}
\vec{\gamma}^\alpha=\left( \begin{array}{lcr}
\cos\phi^\alpha\sin\theta^\alpha&\sin\theta^\alpha\sin\phi^\alpha&\cos\theta^\alpha
\end{array} \right)^T
\label{eq0}
\end{equation}

Si on utilise des hypothèses de déformation et de champ homogène sur le cristal \cite{daniel}, l'énergie libre totale par domaine ${W}^\alpha$ est la somme de trois contributions énergétiques:

\begin{itemize}
 \item  L'énergie de champ: ${W}_H^\alpha=-\mu_0\vec{H}.\vec{M}^\alpha$.\\
 \item  L'énergie magnéto-mécanique: ${W}_\sigma^\alpha=- \boldsymbol{\sigma}:\boldsymbol{\epsilon}^\alpha_\mu$.\\
 \item  L'energie magnétocristalline: ${W}_K^\alpha=K_1\left(\gamma_1^2 \gamma_2^2 + \gamma_2^2\gamma_3^2 + \gamma_1^2\gamma_3^2 \right)$
\end{itemize}

$\mu_0$ est la perméabilité du vide ($4\pi.10^{-7}$ H/m), $\vec{H}$ et $\boldsymbol{\sigma}$ sont le champ magnétique et  le tenseur des contraintes appliqués au cristal, $\boldsymbol{\epsilon}^\alpha_\mu$ est le tenseur de magnétostriction du domaine considéré (\ref{epsmu}), $K_1$ est la constante d'anisotropie principale du matériau.

\begin{equation}
\boldsymbol{\epsilon}^\alpha_\mu=\frac{3}{2}\left( \begin{array}{ccc}
 \lambda_{100}(\gamma_1^2 - \frac{1}{3}) & \lambda_{111}(\gamma_1\gamma_2) &\lambda_{111}(\gamma_1\gamma_3) \\
  & \lambda_{100}(\gamma_2^2 - \frac{1}{3}) & \lambda_{111}(\gamma_2\gamma_3) \\
sym & &\lambda_{100}(\gamma_3^2 - \frac{1}{3}) \\
\end{array} \right)
\label{epsmu}
\end{equation}

\begin{equation}
\vec{n_c}=\left(\begin{array}{lcr}
\cos\phi_c\sin\theta_c&\sin\theta_c\sin\phi_c&\cos\theta_c
\end{array} \right)^T
\label{dirnc}
\end{equation}

Compte tenu de la symétrie cubique du cristal, une direction de chargement ( champ magnétique ou traction uniaxiale) trouve son équivalent dans le triangle standard défini par les axes $[100]$, $[110]$ et $[111]$ du cristal (figure \ref{mmd}b). On considère une direction de chargement $\vec{n}_c$ définie par les deux angles sphériques $\phi_c$ et $\theta_c$ (\ref{dirnc}). Les directions de chacun des domaines minimisent par définition l'énergie libre totale. Compte tenu de la restriction au triangle standard, il est possible de réaliser une minimisation analytique. On aboutit par exemple aux expressions suivantes pour le domaine (1) de la figure \ref{mmd}a :

\begin{equation}
\begin{array}{l}
\displaystyle{\phi^1(H,\sigma,\vec{n}_c)= \frac{\mu_o M_s H \phi_c + \arctan\left(\frac{3}{2}\lambda_{111}\sigma\sin(2\phi_c)\right)}{\mu_oM_sH + 2K_1 + 3\lambda_{100}\sigma \cos(2\phi_c)}}\\ \\
\displaystyle{\theta^1 (H,\sigma,\vec{n}_c)=\frac{\pi}{2}- \frac{\mu_o M_s H (\frac{\pi}{2}-\theta_c)+ \arctan\left(\frac{3}{2}\lambda_{111}\sigma\sin(2(\frac{\pi}{2}-\theta_c)\right) }{\mu_oM_sH + 2K_1 + 3\lambda_{100}\sigma \cos(2(\frac{\pi}{2}-\theta_c))}}
\end{array}
\end{equation}
avec $\vec{H}=H.\vec{n}_c$ et $\vec{\sigma}=\sigma.\vec{n}_c$ les vecteurs champ magnétique et contrainte. Les fractions de chaque domaine $f^\alpha$ sont ensuite obtenues en utilisant une relation explicite inspirée d'une fonction statistique de Boltzmann \cite{daniel}:
\begin{equation}
f^\alpha=\frac{exp(-A_s.{W}^\alpha)}{\sum_\alpha exp(-A_s.{W}^\alpha)}
\end{equation}
$A_s$ est un paramètre proportionnel à la susceptibilité initiale $\chi_{0}$ du matériau tel que $A_s=\frac{3\chi_{0}}{\mu_0M_s^2}$. Le comportement étant considéré homogène dans le monocristal, les règles d'homogénéisation s'appliquent pour le calcul des valeurs moyennes de l'aimantation et la déformation: 
\begin{equation}
\left\{\begin{array}{l}
\vec{M}= \sum_\alpha f^\alpha \vec{M}^\alpha \\
\epsilon^\mu=\sum_\alpha f^\alpha \epsilon_\mu^\alpha 
\end{array}\right .
\end{equation}

On aboutit à un modèle capable de rendre compte du comportement magnéto-mécanique d'un monocristal. Or, puisque toutes les directions de chargement possibles peuvent être restreintes au triangle standard, le comportement d'un polycristal isotrope est nécessairement donné par un chargement le long d'une direction appartenant à ce triangle.  La non linéarité des phénomènes en présence implique que cette direction n'est pas une direction moyenne, et est susceptible de changer d'orientation avec la contrainte ou le niveau du champ magnétique imposé. On fait l'hypothèse que ce changement est assez petit pour être négligé \cite{lazreg2}. La direction moyenne théorique est obtenue pour $\phi_c=39^o$ et $\theta_c=78^o$ \cite{lazreg3}
.


\section{Modélisation d'un acier DP, localisation des champs}

La présence de plusieurs phases de nature différente crée une perturbation locale du champ (magnétique ou mécanique) imposé à chacune des phases. Le champ local ne vaut ainsi généralement pas le champ moyen. On parle de localisation du champ macroscopique considéré. On considère le milieu biphasé ($f$,$m$). On note $I$ une inclusion, jouée alternativement par la phase $f$ ou la phase $m$.
%
	 

Le champ magnétique local $\vec{H}_I$ vu par la phase $I$ est une fonction complexe du champ macroscopique $\vec{H}$ et des propriétés du milieu moyen. Dans le cas d'un problème d'inclusion sphérique \cite{daniel}, et en utilisant une  hypothèse de champ homogène par phase et de comportement linéaire, la relation de localisation prend la forme:

\begin{equation}
\vec{H}_I=\vec{H} + \frac{1}{3 + 2\chi_o}\left(\vec{M}- \vec{M}_I \right)
\end{equation}

$\chi_{0}$ et $\vec{M}$ représentent la susceptibilité et l'aimantation du milieu moyen, $\vec{M}_I $ étant l'aimantation de l'inclusion. L'extension au comportement non linéaire suppose d'utiliser une susceptibilité sécante pour la définition de $\chi_{0}$, soit $\chi_{0}={\|\vec{M} \|}/{\|\vec{H} \|}$. Cette approche est appliquée à la microstructure biphasée ($f$,$m$). On obtient:

\begin{equation}
\begin{array}{lcr}
\vec{H}_f=\vec{H} + \frac{1}{3 + 2\chi_o}\left(\vec{M}- \vec{M}_f \right) &;&
\vec{H}_m=\vec{H} + \frac{1}{3 + 2\chi_o}\left(\vec{M}- \vec{M}_m \right)
\end{array}
\end{equation}

avec:

\begin{equation}
\begin{array}{lcl}
\vec{H}=f_f\vec{H}_f + f_m\vec{H}_m &;& \vec{M}=f_f\vec{M}_f + f_m\vec{M}_m
\end{array}
\end{equation}

Les champs $\vec{H}_f$ et $\vec{H}_m$ sont introduits en entrée de deux calculs multidomaines. Ces calculs fournissent une aimantation par phase, puis une moyenne permettant une nouvelle évaluation des champs locaux. Ce processus est itéré jusqu'à stabilité des grandeurs magnétiques (illustré figure \ref{conv}a).


La solution du problème d'inclusion d'Eshelby constitue la base de modélisation du comportement des milieux hétérogènes en mécanique. La déformation de magnétostriction $\boldsymbol{\epsilon}_I^\mu$ est la déformation que subirait l'inclusion considérée en l'absence de la résistance exercée par la matrice. Aux contraintes d'incompatibilité s'ajoute une contrainte associée au chargement extérieur $\boldsymbol{\sigma}$. Si de plus on décompose les déformations en une somme de déformation élastique et de déformation de magnétostriction et qu'on considère les constantes d'élasticité homogènes (même constantes élastiques pour les deux phases), on aboutit à la formulation suivante de la contrainte $\boldsymbol{\sigma}_I$ dans l'inclusion $I$:

 \begin{equation}
\boldsymbol{\sigma}_I=\boldsymbol{\sigma}+  \mathbb{C}(\mathbb{I}-\mathbb{S}^E):(\boldsymbol{\epsilon}^\mu- \boldsymbol{\epsilon}^\mu_I)
\end{equation}

$\mathbb{C}$ est le tenseur d'élasticité du milieu; $\mathbb{S}^E$ est le tenseur d'Eshelby, ne dépendant que des paramètres matériau de la matrice et de la forme de l'inclusion. $\boldsymbol{\epsilon}^\mu$ est le tenseur de magnétostriction moyen. Cette approche est appliquée à la microstructure biphasée ($f$,$m$) en considérant des modules et des formes identiques. On obtient:

\begin{equation}
\begin{array}{lcr}
\boldsymbol{\sigma}_f=\boldsymbol{\sigma}+  \mathbb{C}(\mathbb{I}-\mathbb{S}^E):(\boldsymbol{\epsilon}^\mu- \boldsymbol{\epsilon}^\mu_f)\quad\quad
\boldsymbol{\sigma}_m=\boldsymbol{\sigma}+  \mathbb{C}(\mathbb{I}-\mathbb{S}^E):(\boldsymbol{\epsilon}^\mu- \boldsymbol{\epsilon}^\mu_m)
\end{array}
\end{equation}

avec:

\begin{equation}
\begin{array}{lcl}
\boldsymbol{\sigma}=f_f\boldsymbol{\sigma}_f + f_m\boldsymbol{\sigma}_m &;& \boldsymbol{\epsilon}^\mu=f_f\boldsymbol{\epsilon}^\mu_f + f_m\boldsymbol{\epsilon}^\mu_m
\end{array}
\end{equation}

Les champs $\boldsymbol{\sigma}_f $ et $\boldsymbol{\sigma}_m$ sont introduits en entrée de deux calculs multidomaines au même titre que les champs magnétiques. Ces calculs fournissent une déformation et une aimantation par phase, puis des moyennes permettant une nouvelle évaluation des champs mécaniques et magnétiques locaux. Ce processus est itéré jusqu'à stabilité des grandeurs. Restant dans un cadre mécanique linéaire, la mise en place numérique de cette deuxième localisation ne pose pas de problème majeur.

\section{Résultats expérimentaux et modélisation}

\begin{figure}[ht!]
 \centering
{\includegraphics[width=7cm]{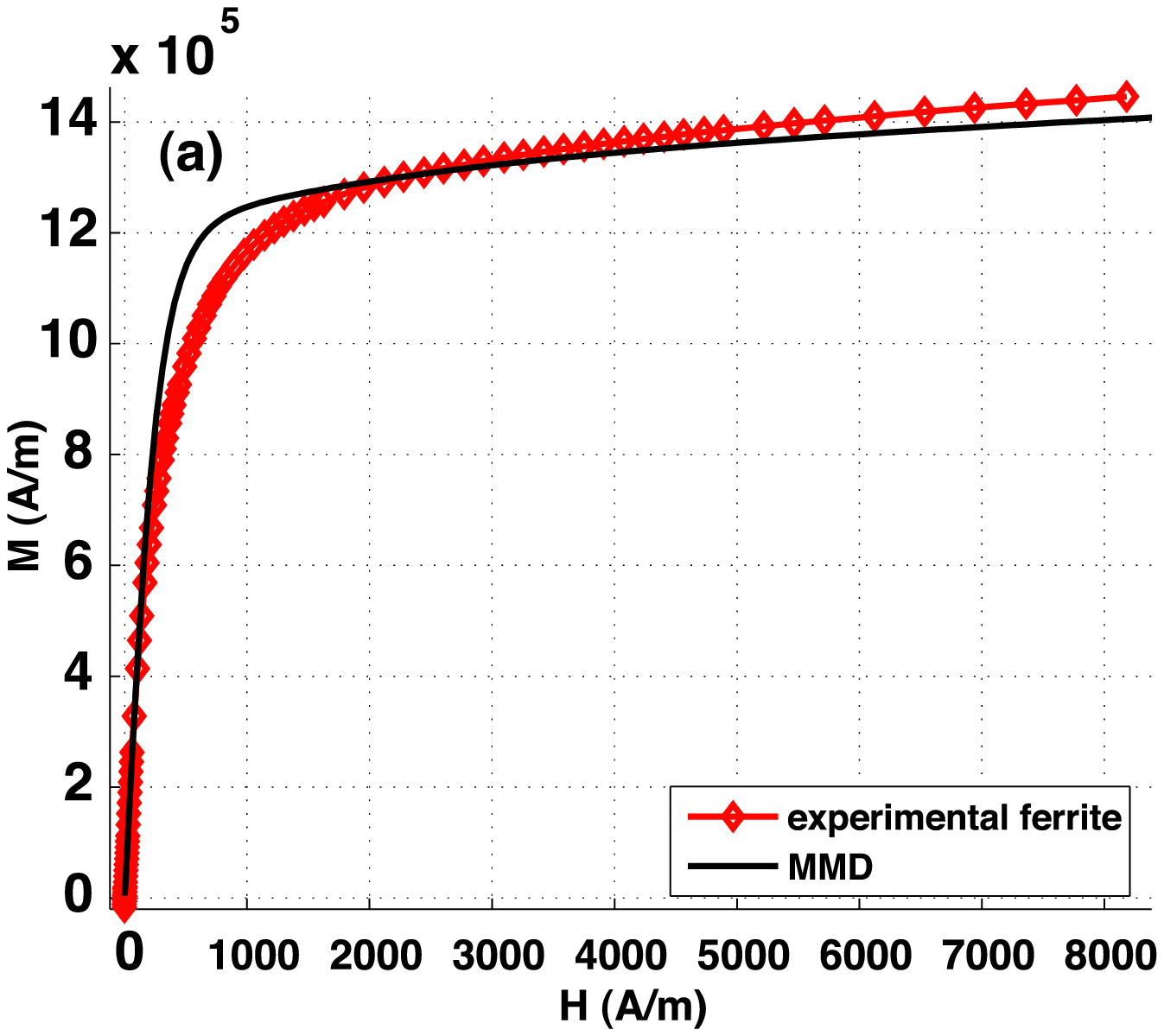}}\quad
{\includegraphics[width=7cm]{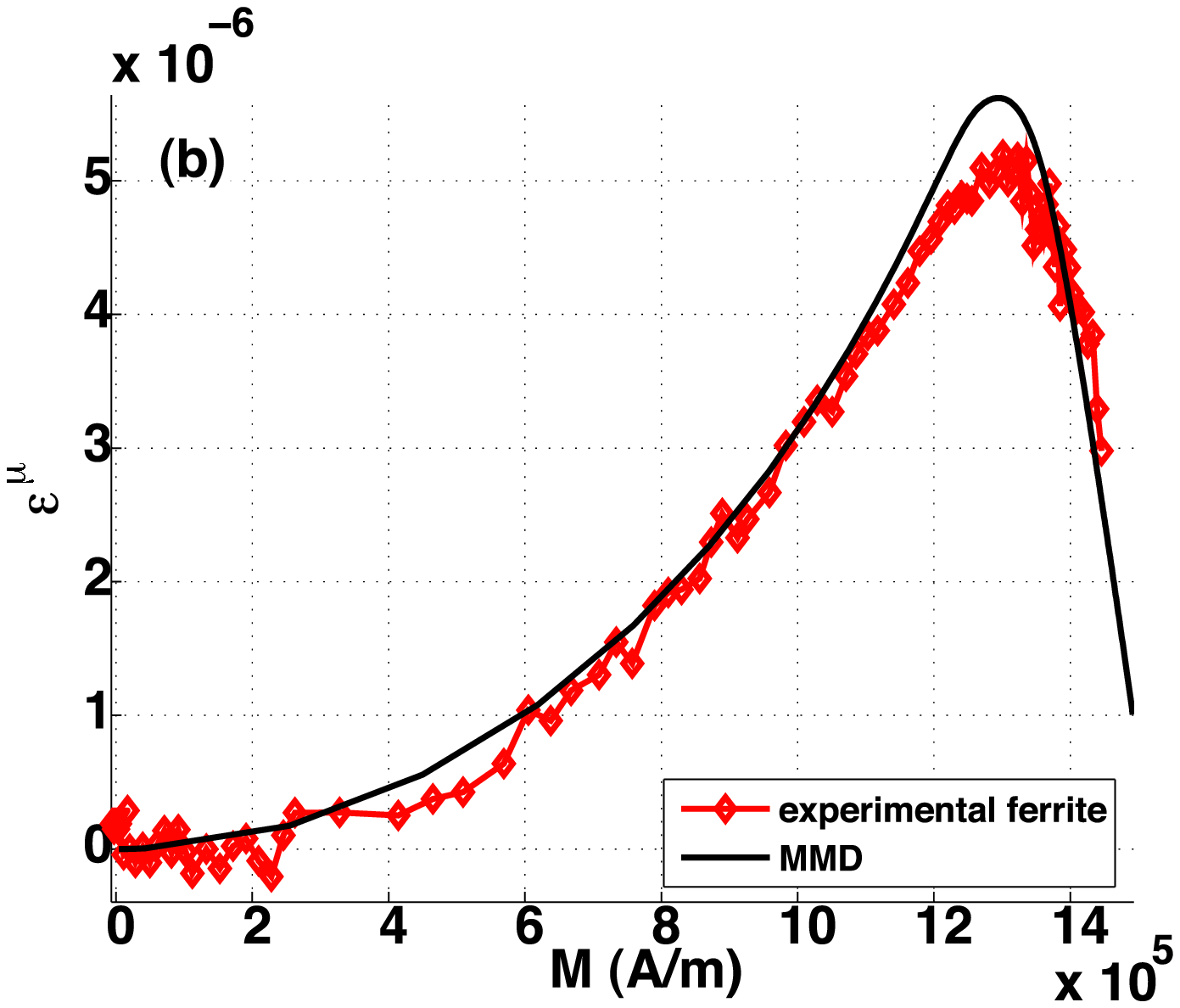}} 
{\includegraphics[width=7cm]{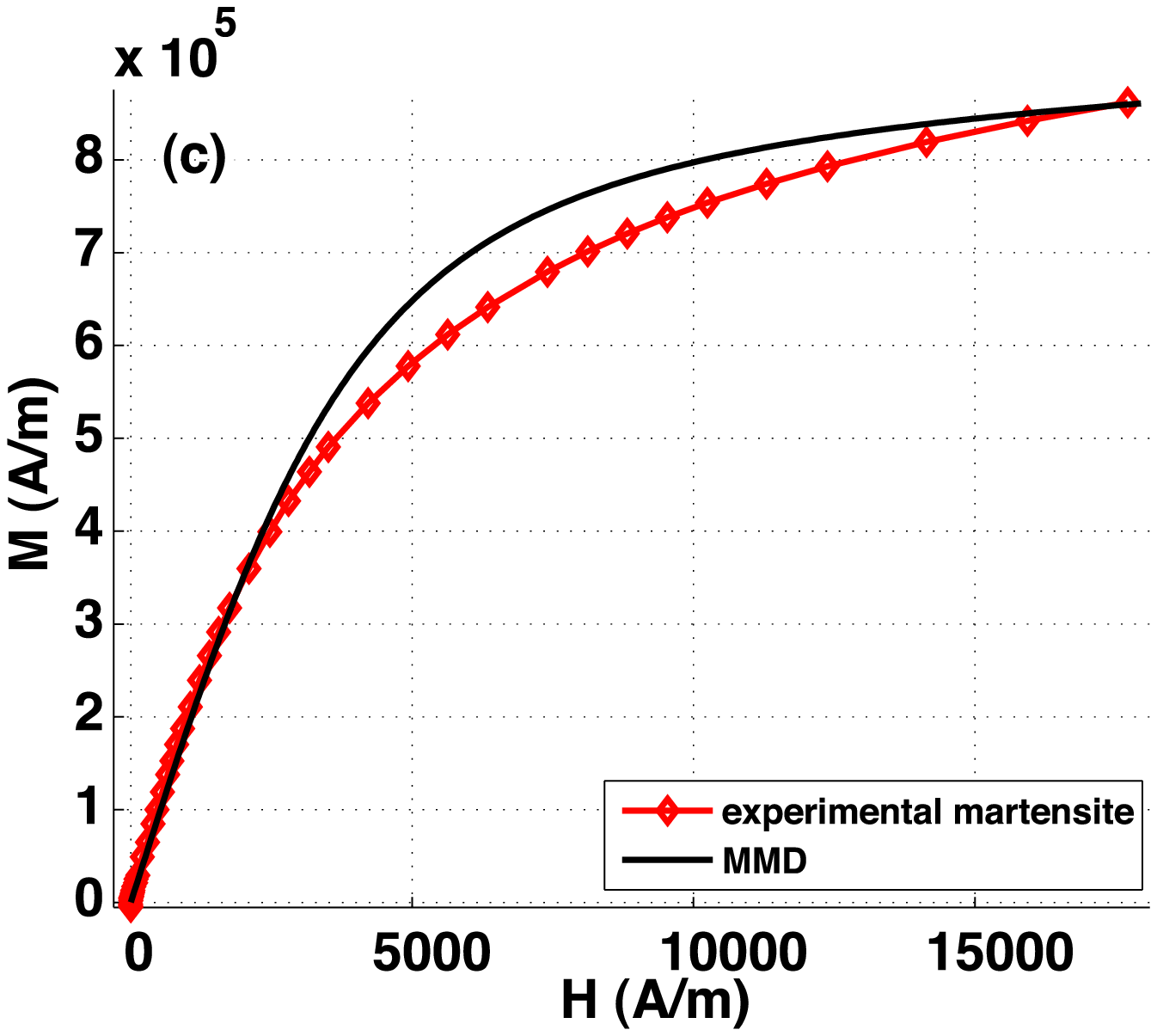}}\quad
{\includegraphics[width=7cm]{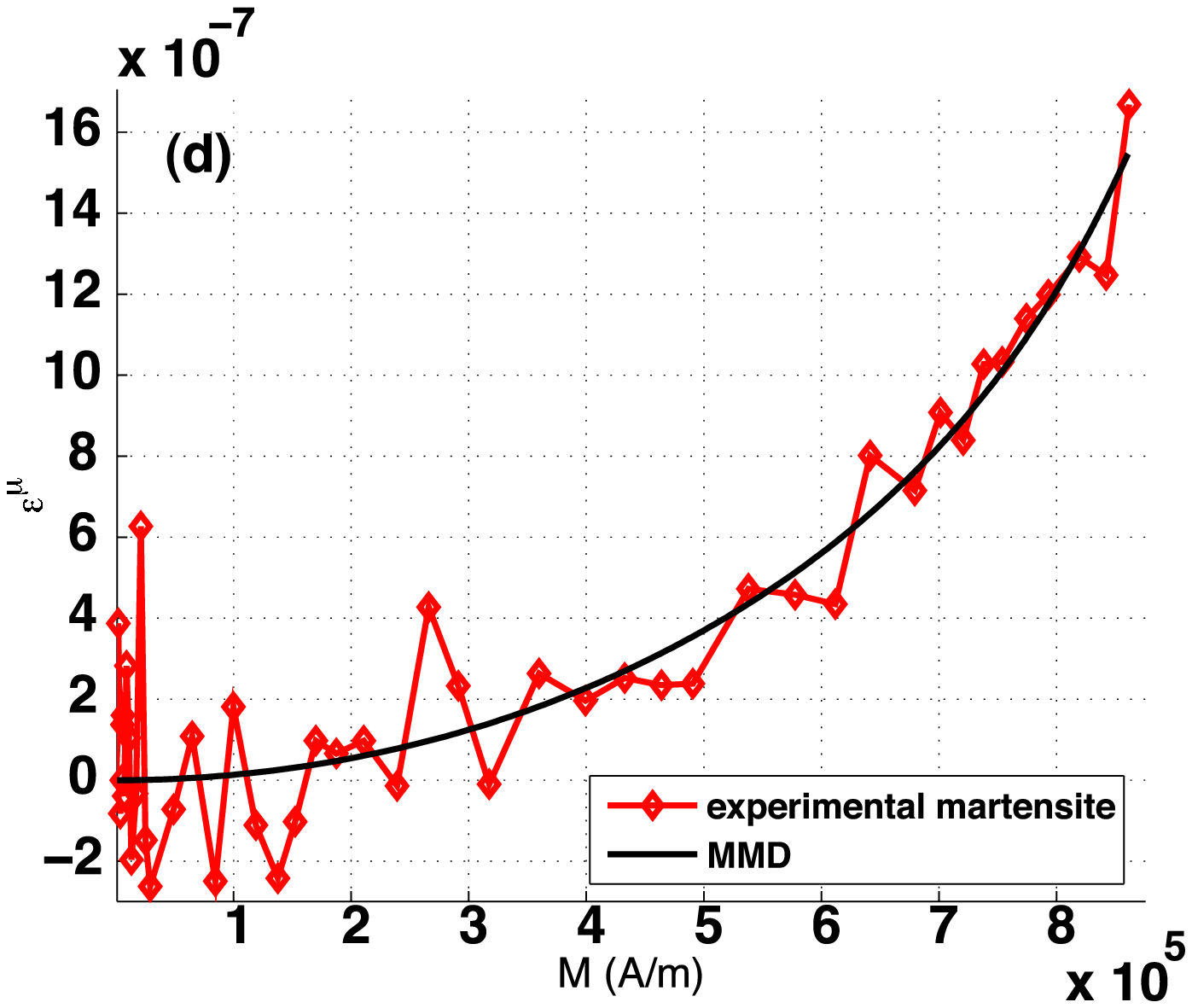}}
\caption{Comportements magnétique (a,c) et magnétostrictif (b,d) du fer pur (a,b) et de la martensite pure (c,d) - comparaisons modèle/expérience.}
\label{expmod} 
\end{figure}

Une première étape a consisté à identifier les paramètres du modèle multidomaine pour les deux phases. Nous disposons d'échantillons de fer pur et d'un acier XC100 trempé à l'eau dont la structure est 100\% martensitique. La figure \ref{expmod} regroupe les résultats expérimentaux obtenus pour les deux matériaux (comportements magnétique et magnétostrictif). On se reportera à \cite{lazreg3} pour les procédures expérimentales. Les simulations à l'aide du modèle multidomaine monophasé sont présentées sur les mêmes figures. Le tableau \ref{table1} regroupe les paramètres utilisés. Ceux du fer sur sont issus de la littérature. Les paramètres de la martensite sont mal documentés. Nous avons procédé à une optimisation pour leur évaluation. Le coude de saturation est relativement mal décrit par le modèle pour les deux matériaux. C’est un défaut associé à l’hypothèse de champ homogène intrinsèque au modèle multidomaine.

%

\begin{table}[ht!]
\centering
\begin{tabular}{lcccccc}
          & $\theta_c(^\circ);\phi_c(^\circ)$&$\lambda_{100}$&$\lambda_{111}$&$K_1(J.m^{-3})$&$M_s(A.m^{-1})$&$A_s(m^3.J^{-1})$\\
  ferrite         &$88; 41$ & $21.10^{-6}$&  $-21.10^{-6}$& $4,8.10^4 $&  $1,71.10^6$ &$3,5.10^{-3}$ \\
  martensite&$90;36$ & $3.10^{-6}$& $3.10^{-6}$& $10.10^4$& $1,05.10^6$ & $4.10^{-4}$ \\
  
  \end{tabular}
\caption{Paramètres utilisés pour la modélisation.}
\label{table1}
\end{table}

\begin{figure}[ht!]
 \centering
{\includegraphics[width=7cm]{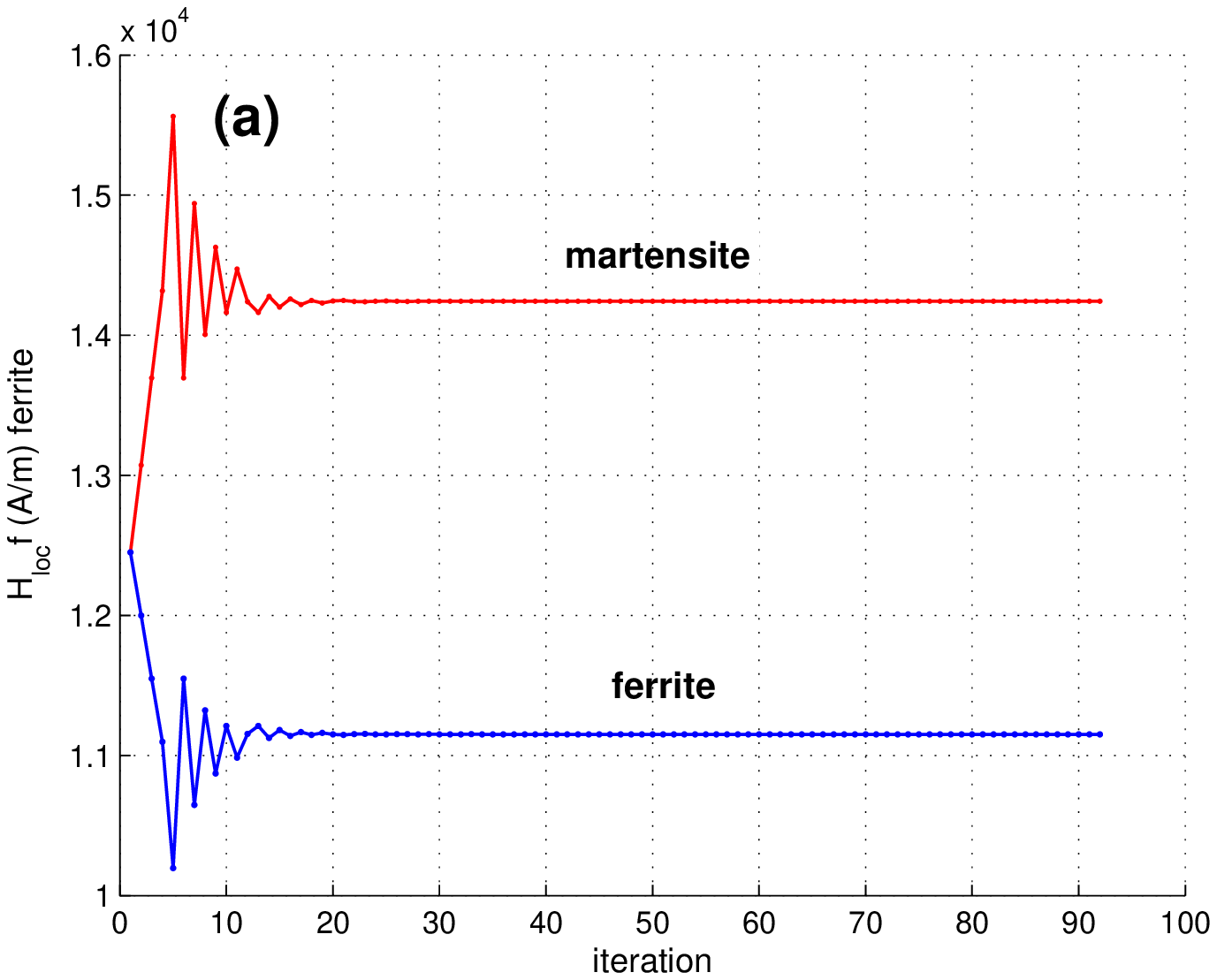}}
{\includegraphics[width=7cm]{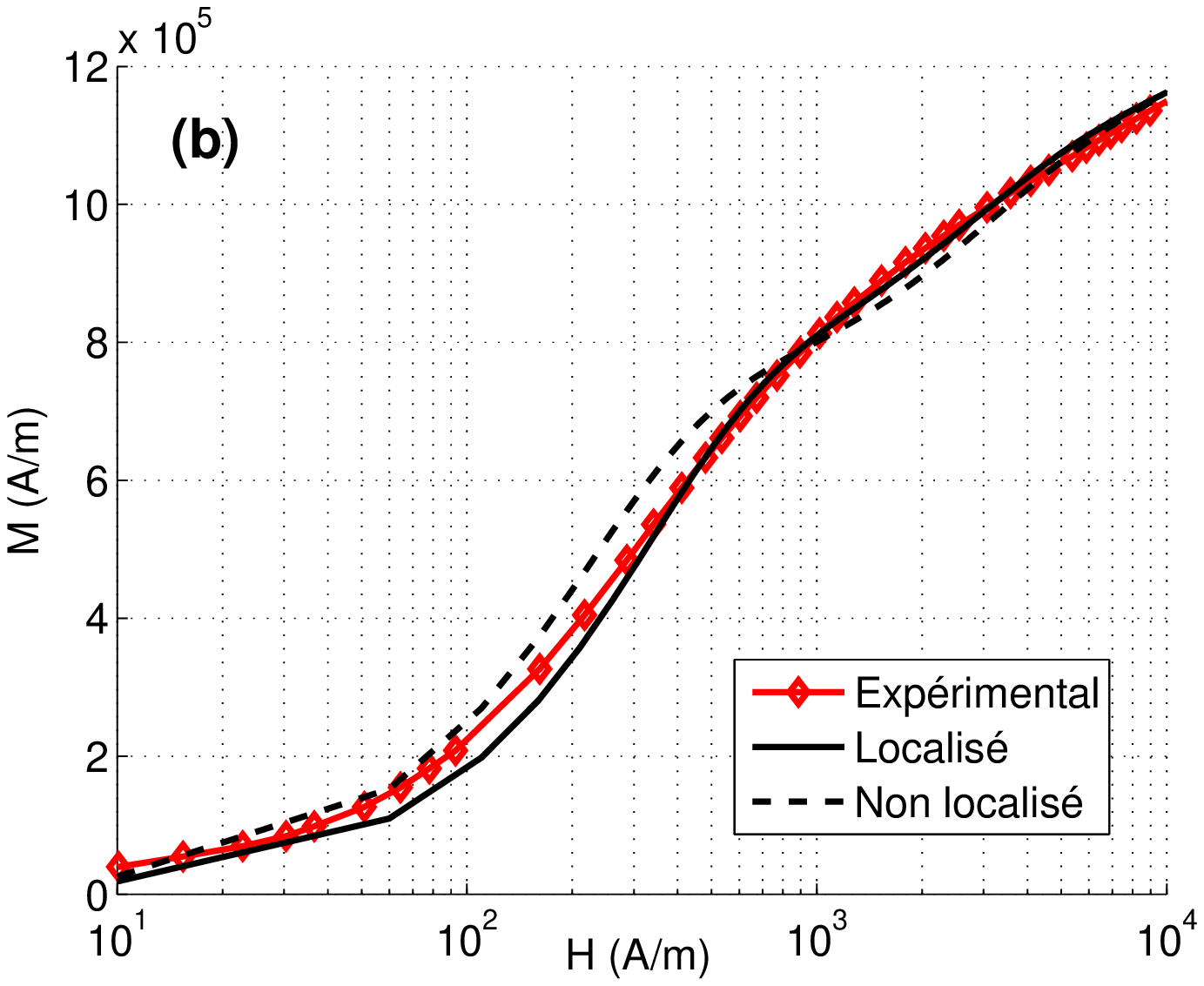}}
\caption{ Illustration de la localisation ($f_m$=42\%) - (a) convergence des champs locaux pour $H=1,25.10^4$ A.m$^{-1}$ -(b) Comportement magnétique: expérience et simulation.}
\label{conv} 
\end{figure}

\begin{figure}[ht!]
 \centering
{\includegraphics[width=7cm]{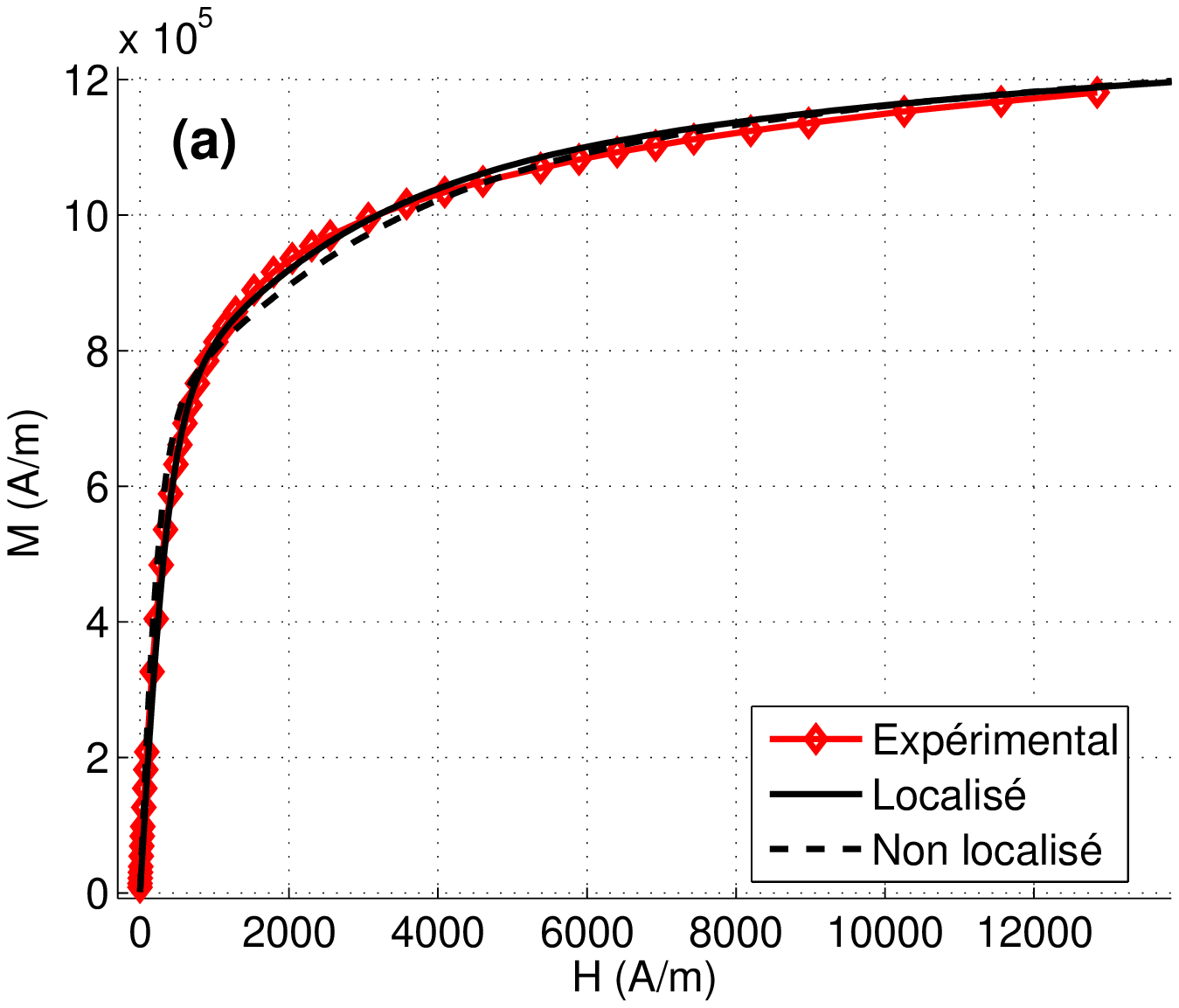}}\quad
{\includegraphics[width=7cm]{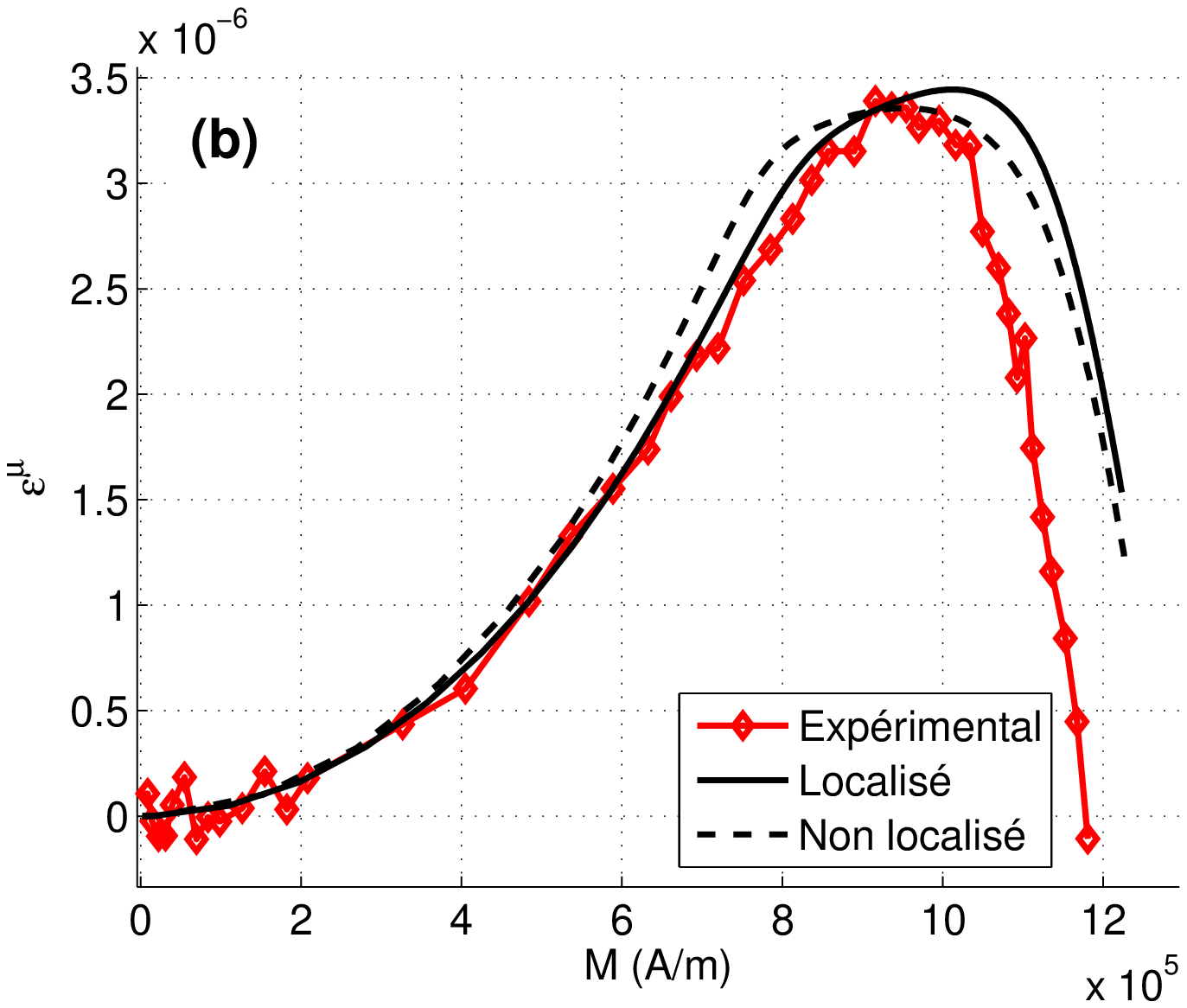}}
\caption{\label{dpLL}Comparaison modèle/expérience pour le comportement magnétique (a) et magnétostrictif (b) du dual-phase ($f_m$=42\%)}
\end{figure}

Une fois les paramètres du comportement de la ferrite et de la martensite identifiés, il est possible de simuler le milieu biphasé. On utilise une fraction de 42\% de martensite (fraction identifiée à partir des micrographies). La figure \ref{conv} permet d'apprécier l'effet de la procédure de localisation en champ (le calcul se fait en contrainte homogène). La figure \ref{conv}a illustre la convergence des champs locaux après quelques itérations: la phase magnétiquement dure (martensite) voit un champ plus élevé que le champ moyen; le champ perçu par la ferrite est plus faible que le champ moyen. La figure \ref{conv}b montre l'effet de cette localisation sur le comportement magnétique. On constate que l'aimantation moyenne prévue est plus faible avec localisation. La figure \ref{dpLL} nous montre le résultat des simulations comparé aux résultats expérimentaux. Comme on peut le voir, le comportement magnétique est bien décrit par le modèle, passé $600A.m^{-1}$, on voit que la procédure de localisation n'a plus qu'un effet marginal. On constate que la déformation de magnétostriction est surestimée   à aimantation élevée. La procédure de localisation en contrainte devrait permettre d'améliorer ce résultat.

\section{Conclusion}


Le modèle proposé permet un estimation rapide du comportement magnéto-mécanique d'une microstructure biphasée. Cette estimation requiert de connaître le comportement de chacune des phases. On constate une surestimation du comportement magnétostrictif, imputable à la non prise en compte de la localisation en contrainte ou à une modélisation trop simpliste (magnétostriction isotrope) de la martensite. La technique développée servira à faire dialoguer des méso-modèles micromagnétiques \cite{ANR}.



\end{document}